\documentclass[aps,showpacs,11pt]{revtex4}
\usepackage{dcolumn}
\usepackage{graphicx}
\usepackage{amsmath}
\usepackage{amsfonts}
\usepackage{amssymb}
\usepackage{psfrag}
\usepackage{wrapfig}
\usepackage{subfigure}
\usepackage{makeidx}
\usepackage{bm}
\usepackage{epsf}
\usepackage{hyperref}

\begin{document}

\title{Holographic parity violating charged fluid dual to Chern-Simons modified gravity}

\author{De-Cheng Zou}
\email{zoudecheng@sjtu.edu.cn}
\author{Yunqi Liu}
\email{liuyunqi@sjtu.edu.cn}
\author{Bin Wang}
\email{wang$_$b@sjtu.edu.cn}

\affiliation{Department of Physics and Astronomy, Shanghai Jiao Tong University, Shanghai 200240, China}


\begin{abstract}
\indent

We discuss the $(2+1)$-dimensional parity violating charged fluid on a finite cutoff surface $\Sigma_c$,
dual to the nondynamical and dynamical Chern-Simons (CS) modified gravities.
Using nonrelativistic long-wavelength expansion method, the field equations are solved up to $\mathcal{O}(\epsilon^2)$
in the nondynamical model. It is shown that there exists nonvortical dual fluid with shear viscosity $\eta$
and Hall viscosity $\eta_A$ on the cutoff surface $\Sigma_c$.
The ratio of shear viscosity over entropy density $\eta/s$ of the fluid takes the universal value $1/{4\pi}$,
while the ratio of Hall viscosity over entropy density $\eta_A/s$ depends on the $\Sigma_c$ and black brane charge $q$.
Moreover the nonvortical dual fluid obeys the magnetohydrodynamic (MHD) equation.
However, these kinematic viscosities $\nu$ and $\nu_A$ related to $\eta$ and $\eta_A$ do not appear in this MHD equation,
due to the constraint condition $\tilde{\partial}^2\beta_j=0$ for the $(2+1)$-dimensional dual fluid.
Then, we extend our discussion to the dynamical CS modified gravity and show that the dual vortical fluid possesses
another so-called Curl viscosity $\zeta_A$, whose ratio to entropy density $\zeta_A/s$ also depends
on the $\Sigma_c$ and $q$. Moreover, the value of $\eta/s$ still equals to $1/4\pi$ and the result of $\eta_A/s$
agrees to the previous result under the probe limit of the pseudo scalar field at the infinite boundary
in the charged black brane background for the dynamical CS modified gravity. This vortical dual fluid
corresponds to the magnetohydrodynamic (MHD) turbulence equation in plasma physics.
\end{abstract}



\maketitle
\section{Introduction}
\label{1s}

Recently there have been a lot of studies on the fluid/gravity dualities \cite{Maldacena:1997re, Gubser:1998bc, Witten:1998qj},
which are considered as special applications of the AdS/CFT correspondence \cite{Bhattacharyya:2008jc,  Rangamani:2009xk, Banerjee:2008th}.
It was argued  that the dual field theory at the AdS boundary can be described by hydrodynamics in the long-wavelength limit.
The ability to derive hydrodynamic equations and transport coefficients from this
duality provides fresh perspectives in understanding holography. A remarkable description of the fluid/gravity duality
was further setup on a finite cutoff surface $\Sigma_c$ outside the horizon \cite{Bredberg:2010ky}.
The discussions have been extended to different models in Einstein
relativity \cite{Bredberg:2011jq, Compere:2011dx, Cai:2012vr,Lysov:2013jsa} and modified gravity models with higher
order curvatures corrections \cite{Niu:2011gu,Cai:2011xv, Zou:2013ix}. Imposing the Petrov-like
condition on the $\Sigma_c(r=r_c)$ in the near horizon limit, the
incompressible Navier-Stokes equations (or modified equations) for a
fluid living on the flat (or spatially curved) spacetime with one
fewer dimensions have been demonstrated in \cite{Lysov:2011xx,
Huang:2011he, Huang:2011kj, Zhang:2012uy,Ling:2013kua, Wu:2013mda,Cai:2013uye}. The physics on a finite
cutoff surface $\Sigma_c$ with finite energy scale is appealing
since it could be reached by experiments.  The study of holography on the finite  surface $\Sigma_c$
may be helpful to understand the microscopic origin of gravity. Other recent works on the
fluid/gravity correspondence can be found
in \cite{Matsuo:2012pi,Berkeley:2012kz, Ashok:2013jda, Green:2013zba, McInnes:2013wba, Adams:2013vsa, Brattan:2013wya}.

Besides the shear viscosity $\eta$ and bulk viscosity $\zeta$ appearing in the usual hydrodynamic system, we know that in
the parity violating hydrodynamic system, there exists other important transport coefficients, the Hall viscosity $\eta_A$
and Curl viscosity $\zeta_A$, which are often studied in the condensed matter physics.
The Hall viscosity which is a non-dissipative viscosity coefficient
does not contribute to the entropy production of the fluid and has been frequently investigated
in field theory approach \cite{Read:2008rn, Kimura:2010yi, Hughes:2011hv, Nicolis:2011ey, Hoyos:2011ez, Bradlyn:2012ea, Hughes:2012vg, Barkeshli:2012}.
In the quantum Hall fluids, at zero temperature, the usual dissipative
shear and bulk viscosities vanish, while the non-dissipative Hall viscosity can be nonzero provided that
the quantum Hall fluid has energy gap and broken time-reversal symmetry \cite{Avron:1995fg}.
How can we study this Hall viscosity from holography?
This is an interesting question to pursue.

Recently, the fluid/gravity duality was explored in system with parity violation. The CS modified gravity generally includes
parity violating gravitational even including electromagnetic CS terms in the action \cite{Jackiw:2003pm, Alexander:2009tp},
which is  considered as a simple model to realize the holographic description of a $(2+1)$-dimensional
isotropic fluid with broken spatial parity \cite{Saremi:2011ab, Chen:2011fs, Jensen:2011xb, Chen:2012ti}.
It is expected  that in this gravity model the dual fluid may possess a non-zero Hall viscosity at the AdS boundary.
Since the Hall viscosity is related to the presence of a non-trivial background scalar field,
it is natural to anticipate that it encodes  the parity violation in CS gravity.
The Hall viscosity of the dual fluid can be affected by the  electromagnetic field
if one considers the influence by the  electromagnetic CS term on the phase transition
of Holographic Superconductors in four dimensions \cite{Tallarita:2010vu, Tallarita:2010uh}.
In addition, the vorticity of holographic fluid is another interesting property and
the holographic boundary vortical fluids to analogue gravity systems
have been explored in \cite{Leigh:2011au, Caldarelli:2012cm, Leigh:2012jv, Eling:2013sna, Mukhopadhyay:2013gja}.

Using the nonrelativistic fluid expansion method, Cai, et.al.,\cite{Cai:2012mg} have investigated the $(2+1)$-dimensional
parity violating hydrodynamics, dual to the dynamical CS modified gravity on a finite cutoff surface $\Sigma_c$
outside the uncharged black brane horizon. They have presented the dual hydrodynamics with Hall viscosity and Curl viscosity
obeying the incompressible Navier-Stokes equations.
Note that the CS gravity model have two frameworks, the dynamical one and the non-dynamical one,
which are classified by whether or not there is a kinetic term for the scalar field in the action \cite{Motohashi:2011ds}.
In this paper, we will extend the study to discuss the holographic hydrodynamics
dual to non-dynamical and dynamical CS modified gravity respectively.
Besides the gravitational CS term,  we will include the electromagnetic CS term in our discussion.
We will show that the holographic fluid/gravity duality can be realized both in the non-dynamical and the dynamical
CS gravities. In the nondynamical model, the dual nonvotical fluid possesses the shear viscosity $\eta$
and Hall viscosity $\eta_A$ and obeys the magnetohydrodynamic (MHD) equation.
However, these kinematic viscosities $\nu$ and $\nu_A$ related to $\eta$ and $\eta_A$ do not appear in this MHD equation,
due to the constraint condition $\tilde{\partial}^2\beta_j=0$. Here the ratio $\eta/s$ of the fluid equals to $1/4\pi$,
while the ratio $\eta_A/s$ depends on the cutoff surface $\Sigma_c$ and black brane charge $q$.
As to the dynamical model, the dual vortical fluid obeys the magnetohydrodynamic (MHD) turbulence equation in plasma physics.
Besides the shear and Hall viscosities, the dual fluid possesses another so-called Curl viscosity $\zeta_A$,
whose ratio to entropy density $\zeta_A/s$ depends on the $\Sigma_c$ and black brane charge $q$.

The outline of this paper is as follows. In Sec.~\ref{2s}, we adopt two finite diffeomorphism transformations and
make non-relativistic hydrodynamic expansion to a general black brane metric, the pseudo scalar and electromagnetic fields.
By applying this formulism to non-dynamical CS modified gravity
coupled to the electromagnetic field, we calculate the stress-energy tensor of the dual fluid through the Brown-York tensor
and analyze the properties of the dual fluid on the cutoff surface $\Sigma_c$.
In Sec.~\ref{3s}, we extend the above investigation to the dynamical CS modified gravity.
We finally summarize our results in Sec.~\ref{4s}.

\section{Dual fluid to non-dynamical CS modified gravity}
\label{2s}

With the electromagnetic CS term $\theta\tilde{F}F$, the action of non-dynamical CS modified gravity model reads \cite{Li:2009rt}
\begin{eqnarray}
{\cal I}_G=\frac{1}{16\pi G}\int{d^4x\sqrt{-g}\left(R-2\Lambda-4\pi GF_{\mu\nu}F^{\mu\nu}\right)}
+\frac{1}{4}\int{d^4x\sqrt{-g}\left(\lambda_1\theta\tilde{R}R+\lambda_2\theta\tilde{F}F\right)},\label{7a}
\end{eqnarray}
where $\lambda_1$ and $\lambda_2$ are the coupling constants,
$\theta\tilde{R}R$ and $\theta\tilde{F}F$ are gravitational and electromagnetic CS terms with
\begin{eqnarray}
\tilde{R}^{~~\rho\tau}_{\mu\nu~~}&=&\frac{1}{2}\epsilon^{\rho\tau\chi\varphi}R_{\mu\nu\chi\varphi},\quad
\tilde{F}^{\mu\nu}=\frac{1}{2}\epsilon^{\mu\nu\rho\tau}F_{\rho\tau},\nonumber\\
\tilde{R}R&=&\tilde{R}^{\mu\nu\rho\tau}R_{\nu\mu\rho\tau},\quad
\tilde{F}F=\tilde{F}^{\mu\nu}F_{\mu\nu}.\nonumber
\end{eqnarray}
Here $\epsilon^{\mu\nu\rho\tau}$ is the four dimensional Levi-Civita tensor in the bulk
with the convention $\epsilon^{r\tau x y}=1/\sqrt{-g}$. The strengths of the gravitational and electromagnetic
CS corrections are controlled by the pseudo scalar field $\theta$. Usually the pseudo scalar field $\theta$
is not a constant, but a function of spacetime, thus serving as a deformation function.
If $\theta=$const, CS modified gravity reduces to the Einstein gravity.
The negative cosmological constant $\Lambda$ equals $-3/l^2$, where $l$  is the AdS radius.
We take $l=1$ in what follows for convenience.

As usual, we obtain the field equations by varying the action with respect to the metric, electromagnetic
and pseudo scalar fields respectively, yielding
\begin{eqnarray}
W_{\mu\nu}&=&R_{\mu\nu}-\frac{1}{2}g_{\mu\nu}R+\Lambda g_{\mu\nu}+16\pi G \lambda_1 C_{\mu\nu}
+8\pi G T_{\mu\nu}^{(A)}=0,\label{8a}\\
W^{\nu}_{(A)}&=&\nabla_\mu F^{\mu\nu}-\lambda_2\partial_\mu\theta\tilde{F}^{\mu\nu}=0, \label{9a}\\
W_{(\theta)}&=&\lambda_1\tilde{R}R+\lambda_2\tilde{F}F=0, \label{10a}
\end{eqnarray}
where the stress-energy tensor of the electromagnetic field $T_{\mu\nu}^{(A)}$ and the so-called Cotton tensor $C_{\mu\nu}$ are
\begin{eqnarray}
T_{\mu\nu}^{(A)}=\frac{1}{4}g_{\mu\nu}F_{\alpha\beta}F^{\alpha\beta}-F_\mu^{~\alpha}F_{\nu \alpha}, \quad
C_{\mu\nu}=\theta_{,\sigma}\epsilon^{\alpha\beta\sigma~}_{~~~(\mu}R_{\nu)\beta;\alpha}
+\theta_{;\sigma\tau}\tilde{R}^{\sigma~\tau~}_{~(\mu~\nu)}.\nonumber
\end{eqnarray}

It is interesting to take the covariant derivative of the equations of motion (EOM) Eq.~(\ref{8a})
\begin{eqnarray}
\nabla^{\mu}\left(R_{\mu\nu}-\frac{1}{2}g_{\mu\nu}R+\Lambda g_{\mu\nu}\right)+16\pi G \lambda_1 \nabla^{\mu}C_{\mu\nu}
+8\pi G \nabla^{\mu}T_{\mu\nu}^{(A)}=0. \label{12a}
\end{eqnarray}
As we known, the Bianchi identity enforces $\nabla^{\mu}\left(R_{\mu\nu}-\frac{1}{2}g_{\mu\nu}R+\Lambda g_{\mu\nu}\right)=0$.
The covariant derivatives of the Cotten tensor $C_{\mu\nu}$ and stress-energy tensor of electromagnetic $T_{\mu\nu}^{(A)}$
satisfy \cite{Li:2009rt, Ahmedov:2010rn}
\begin{eqnarray}
\nabla^{\mu}C_{\mu\nu}=-\frac{1}{8}\partial_{\nu}\theta\tilde{R}R, \quad
\nabla^{\mu}T_{\mu\nu}^{(A)}=-\frac{\lambda_2}{4}\partial_{\nu}\theta\tilde{F}F.\label{13a}
\end{eqnarray}
Since the pseudo scalar field is spacetime coordinate dependent which leads to $\partial_{\nu}\theta\neq0$,
then Eq.~(\ref{12a}) reduces to
\begin{eqnarray}
\lambda_1\tilde{R}R+\lambda_2\tilde{F}F=0,\label{13b}
\end{eqnarray}
which is exactly the pseudo scalar field equation $W_{(\theta)}=0$. Hence the pseudo scalar field
equation is not independent of the EOM Eq.~(\ref{8a}).

Considering the traceless properties of $C_{\mu\nu}$ and stress-energy tensor $T_{\mu\nu}^{(A)}$,
we have
\begin{eqnarray}
W_{\mu\nu}=E_{\mu\nu}+16\lambda_1\pi G C_{\mu\nu}=0, \label{14a}
\end{eqnarray}
where $E_{\mu\nu}=R_{\mu\nu}-\Lambda g_{\mu\nu}+8\pi G T_{\mu\nu}^{(A)}$ and we have used the trace of EOM $R=4\Lambda$.

To study the dynamics of the dual fluid in $(2+1)$-dimensional flat spacetime, we assume the general $(3+1)$-dimensional
black brane metric \cite{Saremi:2011ab}
\begin{eqnarray}
ds^2=-f(r)d\tau^2+2H(r)drd\tau+r^2dx_idx^i, \quad i=1, 2. \label{3a}
\end{eqnarray}
Then, the induced metric on the cutoff surface $\Sigma_c(r=r_c)$
outside the horizon $r_h$ with the intrinsic
coordinates $\tilde{x}^a\sim\left(\tilde{\tau}=\sqrt{f(r_c)}\tau, \tilde{x}^i=r_cx^i\right)$ is
\begin{eqnarray}
ds_{2+1}^2&=&\gamma_{ab}dx^{a}dx^{b}=-f(r_c)d\tau^2+r_c^2dx_idx^i\nonumber\\
&=&-d\tilde{\tau}^2+\delta_{ij}d\tilde{x}^id\tilde{x}^j. \label{4a}
\end{eqnarray}
We require the metric Eq.~(\ref{4a}) flat when perturbing the bulk metric Eq.~(\ref{3a}) and will investigate the
dual fluid living on the $\Sigma_c(r=r_c)$.

Substituting the metric Eq.~(\ref{3a}) into field equation Eq.~(\ref{14a}), we find that the
Cotten tensor $C_{\mu\nu}$ automatically vanishes and $R_{\mu\nu}-\Lambda g_{\mu\nu}$ only depends on $r$.
This leads the electromagnetic tensor $T_{\mu\nu}^{(A)}$ to be only $r$-dependent. In this paper,
we only consider the electric field for the static background solution.
Hence the vector-potential $A$ only depends on $r$, which ensures that $T_{\mu\nu}^{(A)}$ is only $r$-dependent.
We set $A_\mu dx^\mu=A(r, q)d\tau$, where $q$ is related to the charge of black hole and then $A'(r,q)$ is
obviously nonzero. As to the pseudo scalar field $\theta$
for the static and stable background configuration, we consider the pseudo scalar field $\theta$ to be spatial
dependent at first without loss of generality. From the electromagnetic field equation Eq.~(\ref{9a}),
there exists three components of the electromagnetic field equation
\begin{eqnarray}
{W^{x_1}}_{(A)}&=&-\frac{\lambda_2A'(r,q)}{r^2H(r)}\frac{\partial\theta(r,x_1,x_2)}{\partial x_1}=0, \quad
{W^{x_2}}_{(A)}=\frac{\lambda_2A'(r,q)}{r^2H(r)}\frac{\partial\theta(r,x_1,x_2)}{\partial x_2}=0,\label{4b}\\
{W^{\tau}}_{(A)}&=&\frac{1}{rH^3(r)}\left[r H'(r)A'(r,q)-2A'(r)H(r)-r A''(r,q)H(r)\right]=0. \label{18a}
\end{eqnarray}
Therefore the pseudo scalar field $\theta$ only depends on $r$ and is independent of coordinates $x_1$
and $x_2$ to keep Eq.~(\ref{9a}) satisfied for $A'(r,q)\neq0$.
In addition, it is worth noting that the pseudo scalar field $\theta(r)$
for the background solution usually has been employed in some other discussion on the holographic models for
fluid/gravity duality \cite{Saremi:2011ab,Chen:2011fs,Chen:2012ti}.

As to the bulk metric Eq.~(\ref{3a}),
following \cite{Compere:2011dx}, we can introduce two types of diffeomorphism transformations: (i) a Lorentz boost
with constant boost parameter $\beta_i$; (ii) transformation of $r$ and associated rescalings of $\tau$ and $x^i$.
Taking the nonrelativistic hydrodynamic long-wavelength expansion parameterized by $\epsilon\rightarrow 0$, we have
\begin{eqnarray}
\partial_\tau\sim\epsilon^2,\quad \partial_i\sim\epsilon,\quad \partial_r\sim\epsilon^0.\label{5a}
\end{eqnarray}
Together with $\beta^i=\frac{r_c}{\sqrt{f(r_c)}}v^i$, $v_i=v_i(\tau, x^i)$, $P=P(\tau,x^i)$,
and scaling $v_i\sim\epsilon$ and $P\sim\epsilon^2$, we can express
the transformed bulk metric up to $\mathcal{O}(\epsilon^2)$ in the form \cite{Cai:2012mg}
\begin{eqnarray}
ds^2&=&-f(r)d\tau^2+2H(r)drd\tau+r^2dx_idx^i\nonumber\\
    &&-2r^2\left(1-\frac{r_c^2f(r)}{r^2f(r_c)}\right)v_idx^id\tau-\frac{2r_c^2H(r)}{f(r_c)}v_idx^idr\nonumber\\
    &&+r^2\left(1-\frac{r_c^2f(r)}{r^2f(r_c)}\right)\left(v^2d\tau^2
    +\frac{r_c^2v_iv_j}{f(r_c)}dx^idx^j\right)+f(r)\left(\frac{rf'(r)}{f(r)}-\frac{r_cf'(r_c)}{f(r_c)}\right)Pd\tau^2\nonumber\\
    &&+\frac{r_c^2H(r)}{f(r_c)}v^2drd\tau
    +\left(\frac{r_cf'(r_c)H(r)}{f(r_c)}-2H(r)-2rH'(r)\right)P drd\tau\nonumber\\
    &&+\mathcal{O}(\epsilon^3), \label{6a}
\end{eqnarray}
where the terms in last two lines are all of $\mathcal{O}(\epsilon^2)$.

Under these two types of diffeomorphism transformations, both $\theta(r)$ and $A_\mu dx^\mu$ will be expanded.
After promoting $v^i$ and $P$ to be $(\tau,x^i)$-dependent
and adopting the scaling $v_i\sim\epsilon$ and $P\sim\epsilon^2$, we have
\begin{eqnarray}
A(r,q)d\tau &\rightarrow& A(r,q)\left[d\tau-\frac{r_c^2v_i(\tau,x^i)}{f(r_c)}dx^i
+\frac{r_c^2v(\tau,x^i)^2}{2f(r_c)}d\tau\right.\nonumber\\
&&\left.+\left(\frac{r_cf'(r_c)}{2f(r_c)}-\frac{A'(r,q)r}{A(r,q)}\right)P(\tau,x^i) d\tau\right]
+\mathcal{O}(\epsilon^3),\label{15a}\\
\theta(r)&\rightarrow& \theta(r)-r\theta'(r)P(\tau,x^i).\label{16a}
\end{eqnarray}
In this charged configuration, we only focus on the electromagnetic degrees of freedom (DoF) induced by the above
two kinds of diffeomorphisms, which can be roughly regraded as gravitational, and do not turn
on the independent electromagnetic DoF \cite{Niu:2011gu}. Note that the same approach recently has
been adopted to set up a new magnetohydrodynamic/gravity correspondence in higher dimensional
flat Minkowski space for the independent electromagnetic DoF \cite{Lysov:2013jsa}.

Now substituting the perturbed black brane metric Eq.~(\ref{6a}), electromagnetic field Eq.~(\ref{15a}) and pseudo
scalar field Eq.~(\ref{16a}) into field equations Eqs.~(\ref{8a}-\ref{10a}), we have the EOM at $\mathcal{O}(\epsilon^0)$
\begin{eqnarray}
C^{(0)}_{rr}&=&C^{(0)}_{\tau\tau}=C^{(0)}_{ii}=0, \quad E^{(0)}_{rr}=\frac{2H'(r)}{rH(r)}=0,\nonumber\\
E^{(0)}_{\tau\tau}&=&f(r)\left[-3-\frac{f'(r)H'(r)}{2H^3(r)}
+\frac{f''(r)}{2H^2(r)}+\frac{f'(r)}{rH^2(r)}-\frac{4\pi G A'^2(r,q)}{H^2(r)}\right]=0,\nonumber\\
E^{(0)}_{ii}&=&3r^2+\frac{r f(r)H'(r)}{H^3(r)}-\frac{f(r)}{H^2(r)}-\frac{r f'(r)}{H^2(r)}
-\frac{4\pi G r^2A'^2(r,q)}{H^2(r)}=0, \quad (i=1,2), \label{17a}
\end{eqnarray}
the pseudo scalar field equation automatically reaches $W^{(0)}_{(\theta)}=0$ and electromagnetic
field equation ${W^{\tau}}^{(0)}_{(A)}$ takes the same form as Eq.~(\ref{18a}).
From $E^{(0)}_{rr}$, the function $H(r)$ should equal to a constant and we here take $H(r)=1$ for simplify.
Then $f(r)$ and $A(r,q)$ can be obtained in the forms
\begin{eqnarray}
f(r)=r^2-\frac{m}{r}+\frac{q^2}{r^2},\quad A_\mu dx^\mu=\frac{1}{\sqrt{4\pi G}}\frac{q}{r}d\tau.\label{19a}
\end{eqnarray}
The integral constants $m$ and $q$ here are related to the gravitational mass $M=\frac{m V_2}{8\pi G}$ and
the total charge $Q^2=\frac{4\pi q^2}{G}$ respectively.
Moreover $m$ in terms of the real root of $f(r_h)=0$ is $m=r_h^3+\frac{q^2}{r_h}$.
Then the Hawking temperature $T_h$ of the black brane is obtained
\begin{eqnarray}
T_h=\frac{f'(r_h)}{4\pi}=\frac{1}{4\pi}\left(3r_h-\frac{q^2}{r_h^3}\right).\label{20a}
\end{eqnarray}
The condition $3r_h^4\geq q^2$ should be satisfied for $T_h\geq 0$. Moreover the perturbed metric Eq.~(\ref{6a})
also solves these field equations Eqs.~(\ref{8a}-\ref{10a}) at $\mathcal{O}(\epsilon)$.

At $\mathcal{O}(\epsilon^2)$, a correction term
\begin{eqnarray}
ds_c^2=r^2\left(F(r)\sigma_{ij}+F_A(r)\sigma^A_{ij}\right)dx^idx^j, \label{21a}
\end{eqnarray}
needs to be added to the perturbed metric Eq.~(\ref{6a}) to cancel the terms of tensor sector and pseudo scalar sector due to
the spatial $SO(2)$ rotation symmetry of black brane background \cite{Cai:2012mg}. Here
$\sigma_{ij}$ and $\sigma^A_{ij}$ take $\partial_{(i}v_{j)}-\frac{1}{2}\delta_{ij}\partial_{k}v^k$
and $\frac{1}{2}\left(\epsilon_{ik}\sigma_{j}^{~k}+\epsilon_{jk}\sigma_i^{~k}\right)$ respectively.
The gauges $F(r_c)=0$ and $F_A(r_c)=0$ are chosen to keep the induced metric $\gamma_{ab}$ invariant.
Then the equations of pseudo scalar and electromagnetic fields at $\mathcal{O}(\epsilon^2)$ are obtained
\begin{eqnarray}
W^{(2)}_{(\theta)}&=&\left[\frac{\lambda_1 f^2(r)}{r^3H^3(r)}\left(\frac{f'(r)}{f(r)}-\frac{2}{r}\right)^2
+\frac{\lambda_2 (A^2(r,q))'}{2r^2H(r)}\right]\frac{ r_c^2\Omega}{f(r_c)}=0,\label{22a}\\
{W^{r}}^{(2)}_{(A)}&=&\frac{A'(r,q)}{H^2(r)}\partial_iv^i=0,\label{23a}\\
{W^{\tau}}^{(2)}_{(A)}&=&-\frac{F'(r)A'(r,q)}{H^2(r)}\partial_iv^i
+\lambda_2\frac{r_c^2A(r,q)\theta'(r)}{r^2f(r_c)}\Omega=0,\label{24a}
\end{eqnarray}
where Eq.~(\ref{23a}) leads to incompressibility condition $\partial_iv^i=0$ of the
dual fluid on the $\Sigma_c$ for $A'(r,q)\neq0$.
Then we can also obtain the so-called nonvortical condition $\Omega\equiv\epsilon^{ij}\partial_{i}v_{j}=0$
for Eqs.~(\ref{22a})(\ref{24a}) with solutions Eq.~(\ref{19a}). With these incompressibility and
nonvortical conditions, a new constraint condition at $\mathcal{O}(\epsilon^3)$ reads as
\begin{eqnarray}
\partial^2v_i=0,\quad i=1,2.\label{24b}
\end{eqnarray}

From EOM $W_{\mu\nu}^{(2)}=0$, the Cotten tensors $C^{(2)}_{\mu\nu}$ are given by
\begin{eqnarray}
C^{(2)}_{rr}&=&-\frac{r_c^2H(r)}{2r^4f(r_c)}\frac{d}{dr}\left[\frac{r^2f(r)}{H^2(r)}\left(\frac{f'(r)}{f(r)}
-\frac{2}{r}\right)\theta'(r)\right]\Omega,\nonumber\\
C^{(2)}_{\tau\tau}&=&-\frac{r_c^2}{2r^3H(r)f(r_c)}\frac{d}{dr}\left[\frac{rf^{3/2}(r)}{H^2(r)}\left(\frac{f'(r)}{f(r)}
-\frac{2}{r}\right)\theta'(r)\right]\Omega,\nonumber\\
C^{(2)}_{xx}+C^{(2)}_{yy}&=&-\frac{r_c^2}{2H(r)f(r_c)}\frac{d}{dr}\left[\frac{f^{2}(r)}{H^2(r)}\left(\frac{f'(r)}{f(r)}
-\frac{2}{r}\right)\theta'(r)\right]\Omega \label{25a}
\end{eqnarray}
and also vanish with $\Omega=0$. With the help of the incompressibility condition $\partial_iv^i=0$,
$E^{(2)}_{\mu\nu}$ disappears when imposing the requirement
\begin{eqnarray}
\frac{d}{dr}\left[r^2\left(\frac{f(r)}{H(r)}F'(r)+1\right)\right]\tilde{\sigma}_{ij}
+\frac{d}{dr}\left[r^2\left(\frac{f(r)}{H(r)}F'_A(r)
+\frac{\lambda_1 f(r)\theta'(r)}{2H^2(r)}(\frac{f'(r)}{f(r)}
-\frac{2}{r})\right)\right]\tilde{\sigma}^A_{ij}=0. \label{26a}
\end{eqnarray}
Notice that $\tilde{\sigma}_{ij}$ and $\tilde{\sigma}^A_{ij}$ have different tensor structures,
which leads two second order differential equations which can be solved separately as
\begin{eqnarray}
F'(r)&=&\frac{H(r)}{f(r)}\left(\frac{c_F}{r^2}-1\right),\nonumber\\
F'_A(r)&=&\frac{1}{f(r)}\left(\frac{H(r)}{r^2}c_{F_A}-\frac{\lambda_1 f'(r)\theta'(r)}{2H(r)}\right)
+\frac{\lambda_1\theta'(r)}{rH(r)}.\label{27a}
\end{eqnarray}
The integration constants $c_F$ and $c_{F_A}$ are determined by keeping the functions $F(r)$ and $F_A(r)$ regular
at the horizon $r_h$. It is easy to find that the constants $c_F$ and $c_{F_A}$ are
\begin{eqnarray}
c_F=r_h^2, \quad c_{F_A}=\frac{\lambda_1 r_h^2f'(r_h)\theta'(r_h)}{2H^2(r_h)}.\label{28a}
\end{eqnarray}

According to the fluid/gravity duality, the Brown-York tensor on the $\Sigma_c$ can be identified as the
energy-momentum tensor of the dual fluid. Since the existence of the gravitational CS term,
there are three possible contributions that need to be explained: the usual Gibbons-Hawking boundary term,
a term arising from the variation of gravitational CS term and the boundary counterterm.
The Brown-York tensor $T^{BY}_{ab}$ on the $\Sigma_c$ can be derived from
\begin{eqnarray}
T^{BY}_{ab}&=&\frac{1}{8\pi G}\left(K\gamma_{ab}-K_{ab}-T^{cs}_{ab}+\mathcal{C}\gamma_{ab}\right),\label{29a}
\end{eqnarray}
where $\gamma_{ab}=g_{ab}-n_{a}n_b$ is an induced metric on the $\Sigma_c$,
$K$ is the trace of the extrinsic curvature
tensor $K_{ab}$ of $\Sigma_c$ which is defined by $K_{ab}=\gamma^\delta_{~a}\nabla_{\delta}n_b$.
As shown in \cite{Liu:2012zm},
the contribution $T^{cs}_{ab}$ from $\theta\tilde{R}R$ does not contribute to the Brown-York tensor.
$\mathcal{C}$ is an unfixed constant which can bring a finite
result when the cutoff surface goes to the AdS boundary as determined below.

Plugging the perturbed metric Eqs.~(\ref{6a})(\ref{21a}) into Eq.~(\ref{29a}), the Brown-York tensor $T^{BY}_{ab}$
of the dual fluid in the $\tilde{x}^a\sim(\tilde{\tau},\tilde{x}^i)$ coordinates
can be described as
\begin{eqnarray}
\tilde{T}^{BY}_{ab}=\tilde{T}^{(0)}_{ab}+\tilde{T}^{(1)}_{ab}+\tilde{T}^{(2)}_{ab}+\mathcal{O}(\epsilon^3),\label{30a}
\end{eqnarray}
where
\begin{eqnarray}
8\pi G\tilde{T}^{(0)}_{ab}d\tilde{x}^a d\tilde{x}^b&=&-\left(\frac{2\sqrt{f(r_c)}}{r_c}+\mathcal{C}\right)d\tilde{\tau}^2
+\frac{1}{\sqrt{f(r_c)}}\left(\frac{f'(r_c)}{2}+\frac{f(r_c)}{r_c}+\mathcal{C}\right)d\tilde{x}_id\tilde{x}^i,\nonumber\\
8\pi G\tilde{T}^{(1)}_{ab}d\tilde{x}^a d\tilde{x}^b&=&
-\left(\frac{f(r)}{r^2}\right)'_c\frac{r_c^2\beta_i}{\sqrt{f(r_c)}}d\tilde{x}^id\tilde{\tau},\nonumber\\
8\pi G\tilde{T}^{(2)}_{ab}d\tilde{x}^a d\tilde{x}^b&=&\left(\frac{f(r)}{r^2}\right)'_c\frac{r_c^2}{2\sqrt{f(r_c)}}
\left[\left(2P+\beta^2\right)d\tilde{\tau}^2+\left(\beta_i\beta_j
+\kappa P\delta_{ij}\right)d\tilde{x}^id\tilde{x}^j\right]\nonumber\\
&&-\left[\left(1+f(r_c)F'(r_c)\right)\tilde{\sigma}_{ij}
+f(r_c)F'_A(r_c)\tilde{\sigma}_{ij}^A\right]d\tilde{x}^id\tilde{x}^j+\mathcal{O}(\epsilon^3)\label{31a}
\end{eqnarray}
with $\kappa=\frac{r_c^3}{2f(r_c)}\left(\frac{f(r)}{r^2}\right)'_c
-3-r_c\left(\frac{f(r)}{r^2}\right)''_c/\left(\frac{f(r)}{r^2}\right)'_c$.
Here the trace of the stress-energy tensor $\tilde{T}_{ab}$ in
the $\tilde{x}^a\sim(\tilde{\tau},\tilde{x}^i)$ coordinates can be computed up to $\mathcal{O}(\epsilon^2)$ with
$\tilde{T}_c=\tilde{T}^{BY}_{ab}\tilde{\gamma}^{ab}=\frac{2K+3\mathcal{C}}{8\pi G}$.
For the boundary at infinity, we can take the corresponding factor $\mathcal{C}=-2$
to remove the divergence in the energy-momentum tensor.

In these $(2+1)$-dimensional parity violating hydrodynamic systems, the energy-momentum tensor of the fluid
with the first order gradient expansion usually takes the following form
\begin{eqnarray}
\tilde{T}^{ab}=\rho\tilde{u}^{a}\tilde{u}^{b}+p\tilde{ P}^{ab}-2\eta\tilde{\sigma}^{ab}
-\zeta\tilde{\Theta}\tilde{P}^{ab}-2\eta_A\tilde{\sigma}^{ab}_A-\zeta_A\tilde{\Omega}\tilde{P}^{ab},\label{1a}
\end{eqnarray}
where $\tilde{P}_{ab}=\tilde{\gamma}_{ab}+\tilde{u}_{a}\tilde{u}_{b}$. The shear viscosity $\eta$
and the bulk viscosity $\zeta$ are canonical transport coefficients, while the Hall
viscosity $\eta_A$ and curl viscosity $\zeta_A$ arise from the parity violating effect.
Here $\tilde{u}^{a}=\frac{(1,\beta^i)}{\sqrt{1-\beta^2}}$, $\rho$ is the energy density,
$p$ is the pressure, $\tilde{\sigma}_{ab}$ is the shear
and $\tilde{\Theta}=\tilde{\partial}_{a}\tilde{u}^{a}$ describes the expansion.

Under the nonrelativistic long-wavelength expansion, in the above
stress-energy tensor we have $\tilde{\Theta}=0$ by using
the incompressibility condition $\tilde{\partial}_{a}\tilde{u}^{a}\sim\tilde{\partial}_i\beta^i=0$ at the order $\epsilon^2$,
which results in the vanish of the term $\zeta\tilde{\Theta}\tilde{P}_{ab}$.
With the incompressible and nonvortical conditions ($\tilde{\Theta}=0$
and $\tilde{\Omega}=0$), up to $\mathcal{O}(\epsilon^2)$,
the components of the energy-momentum tensor in the non-relativistic limit are given by
\begin{eqnarray}
\tilde{T}_{\tau\tau}&=&\rho+\left(p+\rho\right)\beta^2,\quad
\tilde{T}_{\tau i}=-\left(p+\rho\right)\beta_i,\nonumber\\
\tilde{T}_{ij}&=&\left(p+\rho\right)\beta_i\beta_j+p\delta_{ij}
-2\eta\tilde{\sigma}_{ij}-2\eta_A\tilde{\sigma}^A_{ij}.\label{32a}
\end{eqnarray}
The energy density $\rho_0$ and pressure $p_0$ of the dual fluid at $\mathcal{O}(\epsilon^0)$
take the following form
\begin{eqnarray}
\rho_0&=&-\frac{\sqrt{f(r_c)}}{4\pi Gr_c}-\frac{\mathcal{C}}{8\pi G}, \quad
p_0=\frac{1}{8\pi G\sqrt{f(r_c)}}\left(\frac{f'(r_c)}{2}+\frac{f(r_c)}{r_c}\right)+\frac{\mathcal{C}}{8\pi G},\nonumber\\
\omega&=&\rho_0+p_0=\frac{r_c^2}{16\pi G\sqrt{f(r_c)}}\left(\frac{f(r)}{r^2}\right)'_c. \label{33a}
\end{eqnarray}
Up to $\mathcal{O}(\epsilon^2)$, the energy density $\rho_c$ and the pressure $p_c$ are corrected to be
\begin{eqnarray}
\rho_c=\rho_0+2\omega P, \quad p_c=p_0+\omega\kappa P \label{34a}
\end{eqnarray}
and the transport coefficients such as the shear viscosity $\eta$ and Hall viscosity $\eta_A$ of the dual fluid are given by
\begin{eqnarray}
\eta=\frac{1+f(r_c)F'(r_c)}{16\pi G},\quad \eta_A=\frac{f(r_c)F'_A(r_c)}{16\pi G}. \label{35a}
\end{eqnarray}
Based on Eq.(\ref{28a}), the shear viscosity $\eta$ and Hall viscosity $\eta_A$ are obtained
\begin{eqnarray}
\eta&=&\frac{1}{16\pi G}\frac{r_h^2}{r_c^2}, \quad \eta_A=\frac{\lambda_1}{32\pi G r_c^2}\left(r_h^2\theta'(r_h)f'(r_h)
-r_c^2f'(r_c)\theta'(r_c)+2r_cf(r_c)\theta'(r_c)\right).\label{36a}
\end{eqnarray}
It is worth noting that the Hall viscosity $\eta_A$ depends on the gravitational CS term $\lambda_1\theta\tilde{R}R$.
If taking the vanishing of $\lambda_1\theta\tilde{R}R$ for the parameter $\lambda_1=0$, we have $\eta_A=0$.

From the metric Eq.~(\ref{3a}), we consider a quotient under shift of $x^i$, $x^i\sim x^i+n^i$ with $n^i\in Z$.
The spatial $R^2$ on the $\Sigma_c$ turns out to be an 2-tours $T^2$ with $r_c$-dependent volume $V_2(r_c)=r_c^2$.
Then the entropy density $s_c$ on the $\Sigma_c$ is described by $S/{V_2(r_c)}$ in the
form $\frac{1}{4G}\frac{r_h^{2}}{r_c^{2}}$ \cite{Bredberg:2010ky}.
So, the ratios of shear viscosity and Hall viscosity to entropy density read as
\begin{eqnarray}
\frac{\eta}{s_c}&=&\frac{1}{4\pi},\quad \frac{\eta_A}{s_c}
=\frac{\lambda_1}{8\pi}\left[\left(3r_h-\frac{q^2}{r_h^3}\right)\theta'(r_h)
-\left(3r_h+\frac{3q^2}{r_h^3}-\frac{4q^2}{r_h^2r_c}\right)\theta'(r_c)\right].\label{37a}
\end{eqnarray}
Apparently the ratio $\eta/s_c$ is independent of $r_c$ and does not receive any influence from the gravitational
and electromagnetic CS terms. However, the ratio $\eta_A/s_c$ is cutoff dependent and background dependent.
If we take the cutoff surface to approach the black brane horizon, $r_c\rightarrow r_h$, $\eta_A/s_c$
vanishes. In the infinite boundary limit $r_c\rightarrow\infty$, if we take the following
assumptions $\theta(r_c)\rightarrow 0$, $\eta_A/s_c$ becomes
$\frac{\eta_A}{s_c}=\frac{\lambda_1\theta'(r_h)}{8\pi}\left(3r_h-\frac{q^2}{r_h^3}\right)$, which is nonnegative for $T_h\geq0$.

The local temperature $T_c$ on the $\Sigma_c$ is identified to the temperature of the dual fluid.
With the Tolman relation, we get the local temperature $T_c$
\begin{eqnarray}
T_c=\frac{T_h}{\sqrt{f(r_c)}}=\frac{1}{4\pi\sqrt{f(r_c)}}\left(3r_h-\frac{q^2}{r_h^3}\right).\label{38a}
\end{eqnarray}
With $T_c=0$, the ratio $\eta_A/s_c$ of the dual fluid disappears in the infinite boundary.
It implies that the parity violating dual fluid corresponds to quantum Hall fluid with time-reversal symmetry.

In addition, we define the chemical potential $\mu_c$ as $\mu_c=\frac{1}{4\pi G\sqrt{f(r_c)}}(\frac{q}{r_h}-\frac{q}{r_c})$
and the charge density $q_c=\frac{q}{V_{2}(r_c)}$ with $\frac{q}{r_c^{2}}$ on the $\Sigma_c$.
Then the thermodynamic relation can be verified
\begin{eqnarray}
\omega-s_cT_c=q_c\mu_c.\label{39a}
\end{eqnarray}

The conservation equations of the Brown-York tensor on the $\Sigma_c$, the so-called momentum constraint, can be deduced
from EOM Eq.~(\ref{8a})
\begin{eqnarray}
-\left(R_{\mu\nu}-\frac{1}{2}g_{\mu\nu}R+\Lambda g_{\mu\nu}\right)n^{\mu}\gamma^{\nu}_{~b}
&=&\left(16\lambda_1\pi G C_{\mu\nu}+8\pi G T_{\mu\nu}^{(A)}\right)n^{\mu}\gamma^{\nu}_{~b}\nonumber\\
&\Longrightarrow&\tilde{\partial}^a\tilde{T}^{BY}_{ab}=T_{\mu b}^{(A)}n^\mu,\label{40a}
\end{eqnarray}
where $n^{\mu}$ is the unit normal vector of $\Sigma_c$ and the Cotton tensor $C_{\mu\nu}$ vanishes since it has no
contribution to the source terms of the momentum constraint up to $\mathcal{O}(\epsilon^3)$. Taking index $b=\tau$,
the temporal component of the momentum constraint at $\mathcal{O}(\epsilon^2)$ reads as
\begin{eqnarray}
&&\tilde{\partial}^a\tilde{T}^{BY}_{a\tau}=T_{\mu \tau}^{(A)}n^\mu=\frac{1}{\sqrt{f(r_c)}}T^r_{~\tau}=0,\nonumber\\
&&\Rightarrow-\left(\frac{f(r)}{r^2}\right)'_c\frac{r_c^2}{\sqrt{f(r_c)}}\tilde{\partial}_i\beta^i=0, \label{41a}
\end{eqnarray}
which leads to the incompressibility condition of the dual fluid $\tilde{\partial}_i\beta^i=0$.

Taking index $b=j$, the spatial component of the momentum constraint at $\mathcal{O}(\epsilon^3)$ is given by
\begin{eqnarray}
&&\tilde{\partial}^a\tilde{T}^{BY}_{a j}=T_{\mu j}^{(A)}n^\mu=F_{ja}J^a,\nonumber\\
&\Rightarrow&\left(\frac{f(r)}{r^2}\right)'_c\frac{r_c^2}{2\sqrt{f(r_c)}}\left(\tilde{\partial}_\tau\beta_j
+\beta^i\tilde{\partial}_i\beta_j+\kappa\tilde{\partial}_j P\right)\nonumber\\
&&-\left[\left(1+f(r_c)F'(r_c)\right)\tilde{\partial}^2\beta_j
+f(r_c)F'_A(r_c)\epsilon^{ij}\tilde{\partial}^2\beta_i\right]=F_{j a}J^a. \label{42a}
\end{eqnarray}

With Eq.~(\ref{24b}), the vanishing of $\partial^2v_j$ implies $\tilde{\partial}^2\beta_j=0$ at $\mathcal{O}(\epsilon^3)$.
Then the momentum constraint reduces to
\begin{eqnarray}
&&\tilde{\partial}_\tau\beta_j+\beta^i\tilde{\partial}_i\beta_j+\tilde{\partial}_jP_r=f_j, \quad (j=1,2),\label{43a}\\
&&\tilde{\partial}_i\beta^i=0, \quad \tilde{\Omega}\equiv\epsilon^{ij}\tilde{\partial}_{i}\beta_{j}=0,\label{44a}
\end{eqnarray}
which corresponds to the magnetohydrodynamic (MHD) equation \cite{Davidson,Roberts}.
Note that the nonvortical dual fluid possesses the shear viscosity $\eta$ and Hall viscosity $\eta_A$,
but these kinematic viscosities $\nu$ and $\nu_A$ related to $\eta$ and $\eta_A$ do not appear in Eq.~(\ref{43a}).
It is special for the (2+1)-dimensional dual fluid. Here the pressure density and external force density read as
\begin{eqnarray}
f_j=\frac{F_{j a}J^a}{r_c\omega},\quad
P_r=\frac{\tilde{p}_c-\tilde{p}_0}{\tilde{\rho}_0+\tilde{p}_0}=\frac{\tilde{p}_c-\tilde{p}_0}{\omega}=\kappa P. \label{45a}
\end{eqnarray}
For this external force density $f_j$, the term $F_{j a}J^a$
consists of $F_{ji}J^i$ and $F_{j\tau}J^\tau$, where $F_{j\tau}J^\tau$ arises from the background electric field,
while $F_{ji}J^i$ corresponds to the Lorentz force due to the magnetic field arising from the perturbation of the
background electric field. Moreover, the current $J^a$ dual to the bulk electromagnetic field is obtained
by $J^a=-n_\mu F^{\mu a}$ on the $\Sigma_c$.
We have $J^\tau=-n_rF^{r\tau}$ at $\mathcal{O}(\epsilon^0)$ and $J^i=-n_rF^{ri}$ at $\mathcal{O}(\epsilon)$.
The partial derivative for boundary current $J^a$ satisfies $\partial_{a}J^a\sim\epsilon^2$.
With the electromagnetic field equation Eq.~(\ref{9a}) and the transformation of $\theta(r)$ Eq.~(\ref{15a}),
there exists a current conservation law $\partial_{a}J^{a}=0$ at $\mathcal{O}(\epsilon^2)$, which is not affected by the
pseudo scalar field. This shows that the conservation law of the boundary current $J^a$  coincides with the
incompressibility condition $\tilde{\partial}_i\beta^i=0$ for the constant dual charge density.

On the other hand, the magnetohydrodynamic (MHD) equation Eq.(\ref{43a}) can be expanded in the form
\begin{eqnarray}
&&\tilde{\partial}_1\tilde{\partial}_\tau\beta_2+\tilde{\partial}_1(\beta^1\tilde{\partial}_1\beta_2)
+\tilde{\partial}_1(\beta^2\tilde{\partial}_2\beta_2)+\tilde{\partial}_1\tilde{\partial}_2 P_r=\partial_1 f_2,\label{45b}\\
&&\tilde{\partial}_2\tilde{\partial}_\tau\beta_1+\tilde{\partial}_2(\beta^1\tilde{\partial}_1\beta_1)
+\tilde{\partial}_2(\beta^2\tilde{\partial}_2\beta_1)
+\tilde{\partial}_2\tilde{\partial}_1 P_r=\partial_2 f_1.\label{45c}
\end{eqnarray}
Considering Eq.(\ref{45b})-Eq.(\ref{45c}), we can obtain
\begin{eqnarray}
&&\tilde{\partial}_\tau\left(\tilde{\partial}_1\beta_2-\tilde{\partial}_2\beta_1\right)
+\tilde{\partial}_1(\beta^1\tilde{\partial}_1\beta_2)-\tilde{\partial}_2(\beta^2\tilde{\partial}_2\beta_1)\nonumber\\
&&+\tilde{\partial}_1(\beta^2\tilde{\partial}_2\beta_2)-\tilde{\partial}_2(\beta^1\tilde{\partial}_1\beta_1)
=\partial_1 f_2-\partial_2 f_1.\label{45d}
\end{eqnarray}
Using the nonvortical condition $ \tilde{\Omega}\equiv\epsilon^{ij}\tilde{\partial}_{i}\beta_{j}=0$ and
the incompressibility condition $\partial_i\beta^i=0$, the above equation leads to
\begin{eqnarray}
\tilde{\partial}_\tau\Omega+\beta^j\tilde{\partial}_j\Omega=\epsilon^{ij}\partial_i f_j=0.\label{45e}
\end{eqnarray}
In non-dynamic case, there is a constraint condition $\epsilon^{ij} \partial_i f_j=0$ for the external force.
But this constraint of the external force is of the order $\epsilon^4$,
while the MHD equation we focused on is of the order $\epsilon^3$.
The constraint conditions we considered in the manuscript, such as the nonvortical
and incompressibility conditions, are of the order not higher than $\epsilon^3$.
In the order of $\epsilon^3$, the above MHD equation for the dual fluid is kept.

We can also try to set up the holographic duality between the nondynamical CS gravity and $(2+1)$-dimensional vortical fluid
in the cutoff flat surface $\Sigma_c$, namely $\Omega\neq 0$. Notice that expressions for $C^{(2)}_{\mu\nu}$ (Eq.~(\ref{25a}))
does not disappear for black brane solutions $f(r)$ and $H(r)$ (Eq.~(\ref{19a})).  As did in \cite{Cai:2012mg},
we introduce some correction terms in the perturbed metric to cancel the residual curl scalar $\Omega$ at $\mathcal{O}(\epsilon^2)$
\begin{eqnarray}
ds^2_{s}=\left(-f(r)k(r)d\tau^2+2H(r)h(r)drd\tau+r^2g(r)dx_idx^i\right)\Omega \label{46a}
\end{eqnarray}
and then the overall perturbed metric with Eqs.(\ref{6a})(\ref{21a}) is given by
\begin{eqnarray}
ds_{o}^2=ds^2+ds_c^2+ds^2_s.\label{47a}
\end{eqnarray}

Inserting this overall perturbed metric into the field equations Eqs.~(\ref{8a}-\ref{10a}),
we find that the pseudo scalar field equation still takes the form of Eq.~(\ref{22a}) and
electromagnetic field equations at $\mathcal{O}(\epsilon^2)$ are changed to be
\begin{eqnarray}
{W^{r}}^{(2)}_{(A)}&=&\frac{A'(r,q)}{H^2(r)}\partial_iv^i
+\frac{3h(r)f(r)A'(r,q)}{2H^3(r)}\left[\frac{f'(r)}{f(r)}-\frac{H'(r)}{H(r)}+\frac{h'(r)}{3h(r)}\right]\Omega=0, \label{48a}\\
{W^{\tau}}^{(2)}_{(A)}&=&-\frac{F'(r)A'(r,q)}{H^2(r)}\partial_iv^i
+\left[g'(r)A'(r,q)+\lambda_2\frac{r_c^2A(r,q)\theta'(r)}{r^2f(r_c)}\right]\Omega=0.\label{49a}
\end{eqnarray}
Consider $A'(r,q)\neq0$, $\Omega\neq 0$ and different structures of $\partial_iv^i$ and $\Omega$
in these electromagnetic field equations, Eqs.~(\ref{48a})(\ref{49a}) lead to the incompressible
condition $\partial_iv^i=0$ and
\begin{eqnarray}
&&\frac{f'(r)}{f(r)}-\frac{H'(r)}{H(r)}+\frac{h'(r)}{3h(r)}=0, \label{50a}\\
&&g'(r)A'(r,q)+\lambda_2\frac{r_c^2A(r,q)\theta'(r)}{r^2f(r_c)}=0. \label{51a}
\end{eqnarray}
In addition, in order to satisfy the pseudo scalar field equation Eq.~(\ref{22a}) with $\Omega\neq 0$, we have
\begin{eqnarray}
\frac{\lambda_1 f^2(r)}{r^3H^3(r)}\left(\frac{f'(r)}{f(r)}-\frac{2}{r}\right)^2
+\frac{\lambda_2 (A^2(r,q))'}{2r^2H(r)}=0.\label{51b}
\end{eqnarray}
One can see that Eq.~(\ref{51b}) does not vanish with $H(r)=1$ and black brane solution $f(r)$ Eq.~(\ref{19a}),
while disappears in the trivial case for the locally pure AdS spacetimes with $f(r)=r^2$ and $A(r,q)=0$.
Therefore this new fluid/gravity duality does not set up in the nondynamical CS modified gravity.

It's interesting to note that the action of dynamical CS modified gravity is related to the kinetic term for the pseudo
scalar field $\theta$ \cite{Alexander:2009tp}, it is expected that this dynamical CS modified
gravity can help to overcome the difficulty. We will discuss this possibility in the next section.

\section{Dual fluid to dynamical CS modified gravity}
\label{3s}

The action of dynamical CS modified gravity coupled to the electromagnetic CS term is \cite{Chen:2011fs}
\begin{eqnarray}
{\cal I}_G&=&\frac{1}{16\pi G}\int{d^4x\sqrt{-g}\left(R-2\Lambda
-4\pi G F_{\mu\nu}F^{\mu\nu}\right)}\nonumber\\
&&+\int{d^4x\sqrt{-g}\left(\frac{\lambda_1}{4}\theta\tilde{R}R+\frac{\lambda_2}{4}\theta\tilde{F}F
-\frac{1}{2}\partial_{\mu}\theta\partial^{\mu}\theta-V(\theta)\right)}\label{52a}.
\end{eqnarray}
The corresponding new equations of motion (NEOM), the electromagnetic and scalar fields equations read
\begin{eqnarray}
\hat{W}_{\mu\nu}&=&R_{\mu\nu}-\frac{1}{2}g_{\mu\nu}R+\Lambda g_{\mu\nu}+16\lambda_1\pi G C_{\mu\nu}
+8\pi G T_{\mu\nu}^{(\theta)}+8\pi G T_{\mu\nu}^{(A)}=0, \label{53a}\\
\hat{W}^{\nu}_{(A)}&=&\nabla_\mu F^{\mu\nu}-\lambda_2\partial_\mu\theta\tilde{F}^{\mu\nu}=0, \label{54a}\\
\hat{W}_{(\theta)}&=&\frac{\lambda_1}{4}\tilde{R}R+\frac{\lambda_2}{4}\tilde{F}F+\square\theta-\frac{dV}{d\theta}=0 \label{55a}
\end{eqnarray}
with the stress-energy tensor of pseudo scalar field
\begin{eqnarray}
T_{\mu\nu}^{(\theta)}=-\partial_\mu\theta\partial_\nu\theta+\frac{1}{2}g_{\mu\nu}(\partial\theta)^2+g_{\mu\nu}V(\theta).\nonumber
\end{eqnarray}
Obviously the new electromagnetic field equation $\hat{W}_{(A)}^{\nu}$ is not influenced by the newly
added dynamical terms of pseudo scalar field and still takes the expression as Eq.~(\ref{9a}).
Similarly, the new pseudo scalar field equation can be also derived by the covariant derivative of NEOM
and the NEOM can also be rewritten as
\begin{eqnarray}
\hat{W}_{\mu\nu}=\hat{E}_{\mu\nu}+16\lambda_1\pi G C_{\mu\nu}=0 \label{56a}
\end{eqnarray}
with $\hat{E}_{\mu\nu}=E_{\mu\nu}
-8\pi G\left(\partial_{\mu}\theta\partial_{\nu}\theta+g_{\mu\nu}V(\theta)\right)$.

Similar to the analysis for the case in the non-dynamical CS modified gravity, we also consider the
vector-potential $A$ only relates to $r$ for the background configuration in this case, and then the
pseudo scalar field $\theta$ should be only related to $r$ from the electromagnetic field equation Eq.~(\ref{54a}).
Then substituting the overall perturbed black brane metric Eq.~(\ref{47a}), perturbed pseudo scalar field Eq.~(\ref{16a})
and electromagnetic field Eq.~(\ref{15a}) into NEOM $\hat{W}_{\mu\nu}$ and new pseudo scalar field
equation $\hat{W}_{(\theta)}$ respectively, the background equations for NEOM at $\mathcal{O}(\epsilon^0)$ are obtained
\begin{eqnarray}
\hat{E}^{(0)}_{rr}&=&E^{(0)}_{rr}-8\pi G\theta'^2(r)=0,\quad
\hat{E}^{(0)}_{\tau\tau}=E^{(0)}_{\tau\tau}+8\pi G f(r)V(\theta)=0, \nonumber\\
\hat{E}^{(0)}_{ii}&=&E^{(0)}_{ii}-8\pi G r^2V(\theta)=0,\quad C^{(0)}_{rr}=C^{(0)}_{\tau\tau}=C^{(0)}_{ii}=0. \label{57a}
\end{eqnarray}
The new pseudo scalar field equation at $\mathcal{O}(\epsilon^0)$ can be worked out as
\begin{eqnarray}
\hat{W}^{(0)}_{(\theta)}=-\frac{dV(\theta)}{d\theta}+\frac{\theta'(r)f(r)}{H^2(r)}\left(\frac{2}{r}-\frac{H'(r)}{H(r)}
+\frac{f'(r)}{f(r)}\right)+\frac{\theta''(r)f(r)}{H^2(r)}=0 \label{58a}
\end{eqnarray}
and the new electromagnetic field only have $\tau$-component and takes the same expression for Eq.~(\ref{18a})
in non-dynamical CS modified gravity.

Unfortunately getting the analytic solutions for functions $f(r)$, $H(r)$
and $A(r,q)$ from field equations Eqs.~(\ref{18a})(\ref{57a})(\ref{58a}) is a hard work.
We can expand the functions $f(r)$, $H(r)$, $A(r,q)$ and $\theta(r)$ with a small parameter $\xi$
\begin{eqnarray}
&&f(r)=f_0(r)+\xi f_1(r)+\ldots, \quad H(r)=H_0(r)+\xi H_1(r)+\ldots,\nonumber\\
&&A(r,q)=A_0(r,q)+\xi A_1(r,q)+\ldots, \quad
\theta(r)\rightarrow\xi\theta(r), \quad V(\theta)\rightarrow\xi^2 V(\theta), \label{58c}
\end{eqnarray}
where $f_0(r)$, $H_0(r)$ and $A_0(r,q)$ can be obtained by solving these field equations Eq.~(\ref{57a})
at $\mathcal{O}(\xi^0)$, which reads as Eq.~(\ref{19a}),
\begin{eqnarray}
&&f_0(r)=r^2-\frac{m}{r}+\frac{q^2}{r^2},\quad  H_0(r)=1, \nonumber\\
&&A_0(r,q)=\frac{1}{\sqrt{4\pi G}}\frac{q}{r}. \label{58f}
\end{eqnarray}
Substituting Eq.(\ref{58c}) into field equations Eqs.~(\ref{18a})(\ref{57a})(\ref{58a}), at $\mathcal{O}(\xi)$, we have
\begin{eqnarray}
f_1(r)=-\frac{4\sqrt{\pi G}q C_1}{r^2}, \quad H_1(r)=0, \quad A_1(r,q)=-\frac{C_2}{r}+C_3 \label{58d}
\end{eqnarray}
and the pseudo scalar $\theta$ obeys the following equation
\begin{eqnarray}
\theta''(r)+\left(\frac{f_0'(r)}{f_0(r)}+\frac{2}{r}\right)\theta'(r)-\frac{dV(\theta)/d\theta}{f_0(r)}=0,\label{58e}
\end{eqnarray}
where $C_1$, $C_2$ and $C_3$ are the integral constants. This scalar field equation for the pseudo scalar
field is the same as the Eq.(17) in \cite{Chen:2011fs,Chen:2012ti}. Specifying the form of the potential,
the solutions of this scalar field equation
and asymptotical behaviors of the pseudo scalar field have been also discussed in \cite{Chen:2011fs,Chen:2012ti}.
With these background equations ${\hat{W}}^{(0)}_{\mu\nu}$, ${W^{\tau}}^{(0)}_{(A)}$ and ${\hat{W}}^{(0)}_{(\theta)}$,
we find that the perturbed metric together with the perturbed electromagnetic and pseudo scalar fields automatically
satisfy the field equations Eqs.~(\ref{53a}-\ref{55a}) at $\mathcal{O}(\epsilon)$.

At $\mathcal{O}(\epsilon^2)$, the new electromagnetic field equations Eq.~(\ref{54a}) still take the forms
of Eqs.~(\ref{48a})(\ref{49a}). Then we can also obtain the incompressibility condition $\partial_iv^i=0$
with $\Omega\neq 0$, $A'(r,q)\neq 0$ and different structures of $\Omega$ and $\partial_iv^i$.
The new pseudo scalar field equation $\hat{W}^{(2)}_{(\theta)}$ reads as
\begin{eqnarray}
\hat{W}^{(2)}_{(\theta)}&=&\left[\frac{\theta'(r)f(r)f^{1/2}(r_c)}{H^2(r)}\left(k'(r)-h'(r)+g'(r)\right)
+f^{1/2}(r_c)\left(k(r)-2h(r)\right)\frac{dV(\theta)}{d\theta}\right.\nonumber\\
&&\left.+\frac{\lambda_1 r_c^2f^2(r)}{r^3H^3(r)f(r_c)}\left(\frac{f'(r)}{f(r)}-\frac{2}{r}\right)^2
+\frac{\lambda_2 r_c^2 (A^2(r,q))'}{2r^2H(r)f(r_c)}\right]\Omega.\label{59a}
\end{eqnarray}
Although existence of the pseudo scalar field
makes it hard to get the solution, it can avoid the trivial solution we met in the uncharged case.
Under the incompressibility condition $\partial_iv^i=0$, the non-vanishing components of $\hat{E}^{(2)}_{\mu\nu}$ are expressed as
\begin{eqnarray}
\hat{E}^{(2)}_{rr}&=&\left[\frac{2h'(r)}{r}+\left(\frac{H'(r)}{H(r)}-\frac{2}{r}\right)g'(r)-g''(r)\right]\Omega,\nonumber\\
\hat{E}^{(2)}_{\tau\tau}&=&\left[(g'(r)-h'(r))f'(r)-k'(r)f(r)\left(\frac{H'(r)}{2H(r)}-\frac{7f'(r)}{4f(r)}-\frac{2}{r}\right)
+f(r)k''(r)\right.\nonumber\\
&&\left.-\left(k(r)-2h(r)\right)f(r)/r\left(\frac{H'(r)}{H(r)}-\frac{3f'(r)}{f(r)}-\frac{2}{r}\right)\right.\nonumber\\
&&\left.+8\pi G A'^2(r,q)\left(k(r)-h(r)\right)\right]\frac{f(r)\Omega}{2H^2(r)},\nonumber\\
\hat{E}^{(2)}_{xx}+\hat{E}^{(2)}_{yy}&=&\left[2(h'(r)-k'(r))+r g'(r)\left(\frac{H'(r)}{H(r)}
-\frac{f'(r)}{f(r)}-\frac{4}{r}\right)\right.\nonumber\\
&&\left.-r g''(r)+2\left(k(r)-2h(r)\right)\left(\frac{H'(r)}{H(r)}-\frac{f'(r)}{f(r)}-\frac{1}{r}\right)\right.\nonumber\\
&&\left.+8\pi G r f(r) A'^2(r,q)\left(g(r)-h(r)\right)\right]\frac{r f(r)\Omega}{H^2(r)}\label{60a},
\end{eqnarray}
with the requirement of Eq.~(\ref{26a}). The components of $C^{(2)}_{\mu\nu}$ are shown in Eq.~(\ref{25a}).
With the Dirichlet boundary condition $k(r_c)=0$ and $g(r_c)=0$, the equations of motion Eqs.~(\ref{25a})(\ref{60a}),
pseudo scalar field equation Eq.~(\ref{59a}) and electromagnetic field equation Eqs.~(\ref{48a})(\ref{49a})
at $\mathcal{O}(\epsilon^2)$ are expected to be formally solved with the background equations.
But here we will not concentrate on finding these solutions.

Plugging the overall perturbed metric Eq.~(\ref{47a}) into the Brown-York tensor $T^{BY}_{ab}$ Eq.~(\ref{29a}) of the dual fluid,
$T^{BY}_{ab}$ in the $\tilde{x}^a\sim(\tilde{\tau},\tilde{x}^i)$ coordinates can be described as
\begin{eqnarray}
\tilde{T}^{BY}_{ab}=\tilde{T}^{(0)}_{ab}+\tilde{T}^{(1)}_{ab}+\tilde{T}^{(2)}_{ab}+\mathcal{O}(\epsilon^3),\label{61a}
\end{eqnarray}
where
\begin{eqnarray}
8\pi G\tilde{T}^{(0)}_{ab}d\tilde{x}^a d\tilde{x}^b&=&-\left[\frac{2\sqrt{f(r_c)}}{r_cH(r_c)}+\mathcal{C}\right]d\tilde{\tau}^2
+\left[\frac{f'(r_c)r_c+2f(r_c)}{2r_cH(r_c)\sqrt{f(r_c)}}+\mathcal{C}\right]d\tilde{x}_id\tilde{x}^i\nonumber\\
8\pi G\tilde{T}^{(1)}_{ab}d\tilde{x}^a d\tilde{x}^b&=&-\left(\frac{f(r)}{r^2}\right)'_c
\frac{r_c^2\beta_i}{H(r_c)\sqrt{f(r_c)}}d\tilde{x}^id\tilde{\tau}\nonumber\\
8\pi G\tilde{T}^{(2)}_{ab}d\tilde{x}^a d\tilde{x}^b&=&\left(\frac{f(r)}{r^2}\right)'_c\frac{r_c^2}{2H(r_c)\sqrt{f(r_c)}}
\left[\left(\left(2-\frac{4f(r_c)H'(r_c)}{r_c^2H(r_c)}\left(\frac{f(r)}{r^2}\right)'^{-1}_c\right)P\right.\right.\nonumber\\
&&\left.\left.+\beta^2-\frac{2f(r_c)^{3/2}\left(h(r_c)
-r_cg'(r_c)\right)}{r_c^3}\left(\frac{f(r)}{r^2}\right)'^{-1}_c\tilde{\Omega}\right)d\tilde{\tau}^2\right.\nonumber\\
&&\left.+\left(\beta_i\beta_j+\kappa(r_c) P\delta_{ij}\right)d\tilde{x}^id\tilde{x}^j\right]\nonumber\\
&&-\left[\left(1+\frac{f(r_c)}{H(r_c)}F'(r_c)\right)\tilde{\sigma}_{ij}
+\frac{f(r_c)}{H(r_c)}F'_A(r_c)\tilde{\sigma}_{ij}^A
+\frac{\varpi(r_c)}{2H(r_c)}\tilde{\Omega}\delta_{ij}\right]d\tilde{x}^id\tilde{x}^j\nonumber\\
&&+\mathcal{O}(\epsilon^3)\label{62a}
\end{eqnarray}
with
\begin{eqnarray}
\kappa(r_c)&=&\frac{r_c^3}{2f(r_c)}\left(\frac{f(r)}{r^2}\right)'_c-3-r_c\left(\frac{f(r)}{r^2}\right)''_c/\left(\frac{f(r)}{r^2}\right)'_c
+\frac{(r^2f(r))'_c}{r^3_c}\frac{H'(r_c)}{H(r_c)}, \nonumber\\
\varpi(r_c)&=&\left(f'(r_c)+2f(r_c)/r_c\right) h(r_c)-2f(r_c)\left(g'(r_c)+k'(r_c)\right).\nonumber
\end{eqnarray}
Here we have used the Dirichlet boundary condition $k(r_c)=0$ and $g(r_c)=0$. We can choose the Landau frame
to make the term of $\tilde{\Omega}$ disappear in Eq.~(\ref{62a}), which is necessary to match the
stress-energy tensor of the fluid as we will discuss below.

Now, the components of the stress-energy tensor in the non-relativistic limit with the incompressible
conditions ($\tilde{\Theta}=0$) up to $\mathcal{O}(\epsilon^2)$ are
\begin{eqnarray}
\tilde{T}_{\tau\tau}&=&\rho+\left(p+\rho\right)\beta^2,\quad
\tilde{T}_{\tau i}=-\left(p+\rho\right)\beta_i,\nonumber\\
\tilde{T}_{ij}&=&\left(p+\rho\right)\beta_i\beta_j+p\delta_{ij}
-2\eta\tilde{\sigma}_{ij}-2\eta_A\tilde{\sigma}^A_{ij}-\zeta_A\tilde{\Omega}\delta_{ij}.\label{63a}
\end{eqnarray}
Then the energy density $\rho_0$ and pressure $p_0$ of the dual fluid at $\mathcal{O}(\epsilon^0)$ satisfy
\begin{eqnarray}
\omega=\rho_0+p_0=\frac{r_c^2}{16\pi G H(r_c)\sqrt{f(r_c)}}\left(\frac{f(r)}{r^2}\right)'_c, \label{64a}
\end{eqnarray}
the transformed energy density $\rho_c$ and the pressure $p_c$ up to $\mathcal{O}(\epsilon^2)$ are corrected to be
\begin{eqnarray}
\rho_c=\rho_0+\left(2-\frac{4f(r_c)H'(r_c)}{r_c^2H(r_c)}\left(\frac{f(r)}{r^2}\right)'^{-1}_c\right)\omega P, \quad
p_c=p_0+\omega\kappa(r_c) P \label{65a}
\end{eqnarray}
and the transport coefficients such as the shear viscosity $\eta$, Hall viscosity $\eta_A$
and Curl viscosity $\zeta_A$ of the dual fluid are given by
\begin{eqnarray}
\eta=\frac{1}{16\pi G}\left(1+\frac{f(r_c)}{H(r_c)}F'(r_c)\right),\quad
\eta_A=\frac{1}{16\pi G}\frac{f(r_c)F'_A(r_c)}{H(r_c)},\quad
\zeta_A=\frac{\varpi(r_c)}{16\pi G H(r_c)}. \label{66a}
\end{eqnarray}
With Eq.(\ref{28a}), the shear viscosity $\eta$ takes the same value in non-dynamical CS case and its
ratio $\eta/s_c$ equals $1/4\pi$, the Hall viscosity $\eta_A$ and its ratio are obtained
\begin{eqnarray}
\eta_A&=&\frac{\lambda_1}{16\pi G}\left(\frac{r_h^2\theta'(r_h)f'(r_h)}{2r_c^2H^2(r_h)}
-\frac{f'(r_c)\theta'(r_c)}{2H^2(r_c)}+\frac{f(r_c)\theta'(r_c)}{r_cH^2(r_c)}\right),\nonumber\\
\frac{\eta_A}{s_c}&=&\frac{\lambda_1}{4\pi}\left(\frac{\theta'(r_h)f'(r_h)}{2H^2(r_h)}
-\frac{r_c^2f'(r_c)\theta'(r_c)}{2r_h^2H^2(r_c)}+\frac{r_cf(r_c)\theta'(r_c)}{r_h^2H^2(r_c)}\right)\label{67a}
\end{eqnarray}
and the ratio of Curl viscosity read as $\zeta_A/s_c=\frac{r_c^2\varpi(r_c)}{4\pi r_h^2H(r_c)}$.
However, the other two ratios $\eta_A/s_c$ and $\zeta_A/s_c$ are cutoff dependent and background dependent.

If we take the cutoff surface to approach the black brane horizon, $r_c\rightarrow r_h$, $\eta_A/s_c$
vanishes while $\zeta_A/s_c$ arrives at a finite value $\frac{f'(r_h) h(r_h)}{4\pi H(r_h)}$.
In the infinite boundary limit $r_c\rightarrow\infty$, if we take the following assumptions
\begin{eqnarray}
&&\frac{f(r_c)}{r_c^2}\rightarrow 1-\mathcal{O}(r_h^3/r^3_c),\quad H(r_c)\rightarrow 1,\nonumber\\
&&\theta(r_c) \rightarrow \mathcal{O}(r_h^m/r^m_c),\label{68a}
\end{eqnarray}
$\eta_A/s_c$ becomes
\begin{eqnarray}
\frac{\eta_A}{s_c}=\frac{\lambda_1}{8\pi}\frac{\theta'(r_h)f'(r_h)}{H^2(r_h)}\label{69a}
\end{eqnarray}
when $m>3$. Note that the ratio $\eta_A/s_c$ in our charged black brane background takes the same form
as in the neutral black brane background \cite{Chen:2011fs}. But in our case the electromagnetic field
influence is imprinted in the metric function. Our result agrees with the previous
result by using the probe limit of pseudo scalar field based on the same action
by including the electromagnetic CS terms \cite{Chen:2012ti}.

The conservation equations of the Brown-York tensor on the $\Sigma_c$, the so-called momentum constraint, can be deduced
from NEOM Eq.~(\ref{53a})
\begin{eqnarray}
-\left(R_{\mu\nu}-\frac{1}{2}g_{\mu\nu}R+\Lambda g_{\mu\nu}\right)n^{\mu}\gamma^{\nu}_{~b}
&=&\left(16\lambda_1\pi G C_{\mu\nu}+8\pi G\gamma T_{\mu\nu}^{(\theta)}
+8\pi G T_{\mu\nu}^{(A)}\right)n^{\mu}\gamma^{\nu}_{~b}\nonumber\\
&\Longrightarrow&\tilde{\partial}^a\tilde{T}^{BY}_{ab}=\left(T_{\mu b}^{(\theta)}+T_{\mu b}^{(A)}\right)n^\mu,\label{70a}
\end{eqnarray}
Taking index $b=\tau$, the temporal component of the momentum constraint at $\mathcal{O}(\epsilon^2)$ reads as
\begin{eqnarray}
\tilde{\partial}^a\tilde{T}^{BY}_{a\tau}&=&\left(T_{\mu \tau}^{(\theta)}+T_{\mu \tau}^{(A)}\right)n^\mu=0,\nonumber\\
&&\Rightarrow-\left(\frac{f(r)}{r^2}\right)'_c\frac{r_c^2}{H(r_c)\sqrt{f(r_c)}}\tilde{\partial}_i\beta^i=0, \label{71a}
\end{eqnarray}
which leads to the incompressibility condition of the dual fluid $\tilde{\partial}_i\beta^i=0$.

Taking index $b=j$, the spatial component of the momentum constraint at $\mathcal{O}(\epsilon^3)$ is given by
\begin{eqnarray}
&&\tilde{\partial}^a\tilde{T}^{BY}_{a j}=\left(T_{\mu j}^{(\theta)}+T_{\mu j}^{(A)}\right)n^\mu,\label{72a}\\
&\Rightarrow&\left(\frac{f(r)}{r^2}\right)'_c\frac{r_c^2}{2H(r_c)\sqrt{f(r_c)}}\left(\tilde{\partial}_\tau\beta_j
+\beta^i\tilde{\partial}_i\beta_j+\kappa(r_c)\tilde{\partial}_j P\right)\nonumber\\
&&-\left[\left(1+\frac{f(r_c)}{H(r_c)}F'(r_c)\right)\tilde{\partial}^2\beta_j
+\frac{f(r_c)}{H(r_c)}F'_A(r_c)\epsilon^{ij}\partial^2\beta_i
+\frac{\varpi(r_c)}{2H(r_c)}\epsilon^{ik}\partial_i\partial_j\beta_k\right]\nonumber\\
&&=\frac{r_c\sqrt{f(r_c)}\theta'^2(r_c)}{H^2(r_c)}\partial_j P+F_{j a}J^a,\label{73a}
\end{eqnarray}
With the momentum constraint, we can obtain the incompressible Navier-Stokes equations
\begin{eqnarray}
\tilde{\partial}_\tau\beta_j+\beta^i\tilde{\partial}_i\beta_j+\tilde{\partial}_jP_r-\nu\tilde{\partial}^2\beta_j
-\nu_{A}\epsilon^{ij}\partial^2\beta_i-\xi_A\epsilon^{ik}\partial_i\partial_j\beta_k=f_j,\quad
\tilde{\partial}_i\beta^i=0, \label{74a}
\end{eqnarray}
which corresponds to the magnetohydrodynamic (MHD) turbulence equation with viscosity in
plasma physics \cite{Biskamp}. Here the external force density reads
$f_j=\frac{\sqrt{f(r_c)}\theta'^2(r_c)}{H^2(r_c)\omega}\partial_j P+\frac{F_{j a}J^a}{r_c\omega}$.
Besides the Lorentz force due to the magnetic field which arises from the perturbation
of the electric field and electric force for the electric field,
it is worth noting that the forcing term $f_j$ is also affected by the pseudo scalar field $\theta$
with $\frac{\sqrt{f(r_c)}\theta'^2(r_c)}{H^2(r_c)\omega}\partial_j P$.
The pressure density $P_r$ equals to $\kappa(r_c) P$ and these kinematic viscosities $\nu$, $\nu_{A}$ and $\xi_A$
are defined by $\nu=\eta/\omega$, $\nu_A=\eta_A/\omega$ and $\xi_A=\zeta_A/\omega$.

\section{closing remarks}
\label{4s}

Based on the static black brane metric, we applied the two finite diffeomorphism transformations
and nonrelativistic long-wavelength expansion to derive the bulk
equations of motion up to $\mathcal{O}(\epsilon^2)$ at an
arbitrary cutoff surface $\Sigma_c$ outside the horizon in
the nondynamical and dynamical CS modified gravities.
In this nondynamical model, the dual nonvortical fluid possesses the shear viscosity $\eta$ and Hall viscosity $\eta_A$.
According to the momentum constraint from the conservation equations of the Brown-York tensor,
the dual nonvortical fluid obeys the magnetohydrodynamic (MHD) equation.
However, these kinematic viscosities $\nu$ and $\nu_A$ related to $\eta$ and $\eta_A$ do not appear in this MHD equation,
which is special for the (2+1)-dimensional dual fluid.
The ratio $\eta/s_c$ equals the universal value $1/4\pi$, while the ratio $\eta_A/s$ depends on
the $r_c$ and black brane charge $q$. In the dynamical framework,
besides the shear viscosity $\eta$ and Hall viscosity $\eta_A$,
the dual fluid possesses another so-called Curl viscosity $\zeta_A$,
whose ratio to entropy density $\zeta_A/s$ also depends on the $\Sigma_c$ and black brane charge $q$.
Moreover the dual vortical fluid obeys the magnetohydrodynamic (MHD) turbulence
equation with external force density influenced by the electromagnetic and pseudo scalar fields.
At the infinite boundary, the ratio $\eta_A/s_c$ agrees to the previous
result by using the probe limit of pseudo scalar field in the charged black brane background.
In addition, even though the electromagnetic field is related to the pseudo scalar field, there exists
the current conservation law $\partial_aJ^a=0$ at the order $\epsilon^2$ in both cases,
which is not affected by the pseudo scalar field.

{\bf Acknowledgments}

This work was supported by the National Natural Science Foundation of China.
D.C.Z are extremely grateful to Xiao-Ning Wu, Rong-Gen Cai, Rui-Hong Yue, Shao-Jun Zhang, Xiao-Mei Kuang
and Cheng-Yong Zhang for useful discussions.

\end{document}